\documentclass[twocolumn]{aastex63}
\usepackage{amsmath}

\submitjournal{ApJ}
\shorttitle{FIP and Inverse FIP Effects in Solar Flares}
\shortauthors{Laming}

\begin{document}

\title{The FIP and Inverse FIP Effects in Solar Flares}

\correspondingauthor{J. Martin Laming}
\email{laming@nrl.navy.mil}

\author[0000-0002-3362-7040]{J. Martin Laming}
\affil{Space Science Division, Code 7684, Naval Research Laboratory, Washington DC 20375, USA}

\begin{abstract}
The Inverse First Ionization Potential (FIP) Effect, the depletion in coronal abundance of elements like Fe, Mg, and Si that are ionized in the solar chromosphere relative to
those that are neutral, has been
identified in several solar flares. We give a more detailed discussion of the mechanism of fractionation by the ponderomotive force associated
with magnetohydrodynamic waves, paying special attention to the conditions in which Inverse FIP fractionation arises in order to better understand its
relation to the usual FIP Effect, i.e. the enhancement of coronal abundance of Fe, Mg, Si, etc. The FIP Effect is generated by parallel propagating Alfv\'en waves, with either photospheric, or more likely coronal, origins. The Inverse FIP Effect arises as upward propagating fast mode waves with an origin 
in the photosphere or below, refract back downwards in the chromosphere where the Alfv\'en speed is increasing with altitude. We give a more 
physically motivated picture of the FIP fractionation, based on the wave refraction around inhomogeneities in the solar atmosphere, and inspired by 
previous discussions of analogous phenomena in the optical trapping of particles by laser beams. We apply these insights to modeling the fractionation
and find good agreement with the observations of \citet{katsuda20} and \citet{dennis15}.

\end{abstract}

\keywords{solar wind -- Sun: abundances -- Sun: chromosphere -- turbulence -- waves}

\section{Introduction} \label{sec:intro}
The elemental composition of various regions of the solar corona and wind can vary from that of the underlying photosphere, an inference
first made by \citet{pottasch63}. The usually observed anomaly is an overabundance in the corona by a factor of 3-4 relative to the photosphere 
of elements with first ionization potential (FIP) below about 10 eV, while elements with higher FIP are relatively unaffected. These low FIP elements,
like Mg, Si, Fe, are those than can be photoionized by H I Lyman $\alpha$, and are predominantly ionized in the chromosphere, while the  high FIP
elements (e.g. O, Ne, Ar) remain neutral. A mechanism involving ion-neutral separation in the solar chromosphere is clearly indicated. With the
launch of the Extreme Ultraviolet Explorer (EUVE) satellite in 1992, similar abundance anomalies were also discovered in the corona of late-type stars
\citep[e.g.][]{drake97,laming96,laming99}. The increasing spectroscopic precision afforded by the Chandra and XMM satellites allowed the sample of observed
stellar coronae to increase and a variation of abundance anomaly with stellar spectral type become apparent. As summarized most recently by
\citet{wood18}, stars of spectral type earlier and up to that of the Sun have a similar FIP related coronal abundance anomaly. As the spectral type becomes later, 
the FIP Effect decreases, with coronal abundances becoming essentially photospheric at mid-K. For even later spectral types the abundance anomaly inverts, 
becoming ``Inverse-FIP'', with low FIP elements depleted rather than enhanced in abundance in the corona relative to the photosphere, reaching a 
similar level of depletion, 1/3 - 1/4, to the previously mentioned enhancement.

A similar advance in instrumentation to observe the Sun led to the discovery of Inverse-FIP Effect in solar flares. Using the Hinode/Extreme ultraviolet
Imaging Spectrometer (EIS), \citet{doschek15}  found small patches of enhanced Ar XIV 194.40, 187.96\AA\  
relative to Ca XIV 193.87 \AA\ emission, indicative 
of an Ar/Ca abundance ratio of seven times the photospheric value, in well defined structures near sunspots. \citet{doschek16} also find that S behaves more like a low FIP element in Inverse-FIP Effect, i.e. it is depleted, whereas it is well known to behave as a high FIP element in coronal 
FIP fractionated plasma 
\citep[e.g.][]{laming95}. \citet{doschek17} find regions of reduced FIP (but not actually Inverse FIP) over larger areas. This echoes the work of 
\citet{feldman90}, who found photospheric abundances in the transition region over a sunspot, and argued that the sunspot environment was not
conducive to FIP fractionation. 

The Inverse-FIP effect has also recently been identified in spatially unresolved regions of solar flare plasma, seen in observations with {\it Suzaku} by \citet{katsuda20}.
{\it Suzaku} \citep{mitsuda07} is a Japanese satellite for X-ray astronomy, that due to its low Earth orbit observed the Earth's atmosphere illuminated by the Sun for a few minutes each orbit. \citet{katsuda20} were able to identify Earth albedo emission recorded during four X class flares in 2005 and 2006 with the X-ray Imaging Spectrometer \citep[XIS][]{koyama07}, and after correcting for scattering
and intrinsic albedo emission, isolated the solar emission and measured element abundances. \citet{dennis15} see something similar in 
observations of solar flares by the Solar Assembly for X-Rays \citep[SAX][]{schlemm07} on the MErcury Surface Space ENvironment 
GEochemistry and Ranging (MESSENGER) satellite. They report a transition between fractionated and unfractionated
elements at about a FIP of 7 eV rather than 10 eV (i.e. Ca is fractionated, while other low FIPs are not). We will investigate whether this can come about as a combination of Inverse FIP and FIP Effect, which together deplete low FIP elements like Fe and Si more than Ca.

The FIP fractionation has been identified and modeled in a series of papers \citep{laming04,laming09,laming12,laming15,laming17,laming19} as
being due to the ponderomotive force associated with Alfv\'enic waves. This separates ions from neutrals in the chromosphere, and the sign of the
fractionation depends on the gradient of $\delta E^2/B^2$, the square of the wave electric field divided by the square of the ambient magnetic field. As discussed further below, this result is mathematically clear, but physical picture dictating when this gradient should be positive or negative is more obscure, which limits
our ability to connect abundance observations with other solar phenomena.
Prior work \citep[e.g.][]{laming15} has suggested that parallel propagating waves give FIP, because the degree of reflection required to produce Inverse-FIP proved difficult to achieve in 
Alfv\'en wave propagation calculations, integrating the transport equations in a model solar chromosphere. The total internal reflection of 
initially upward propagating fast mode waves
in the chromosphere with Alfv\'en speed increasing with height appeared to be a plausible Inverse-FIP scenario, investigated further 
observationally by \citet{baker19,baker20}.
The initially upward propagating waves are posited to have an origin in subsurface reconnection as neighboring sunspot umbrae interact.
This seems more likely than the mode conversion of acoustic waves deriving from photospheric convection
in the high plasma $\beta$ photosphere to fast mode waves above the 
$\beta =1$ layer, since the wave amplitudes resulting from such a process are likely insufficient to cause the fractionation.

In this paper we revisit the FIP and Inverse-FIP effects with a much more physically motivated picture than in previous work. Much insight is drawn from
the realization that the ion-neutral separation caused the ponderomotive force is an analog in magnetohydrodynamics (MHD) of phenomena in
optical physics related to the trapping and manipulation of particles in the radiation field of a laser \citep[e.g][]{ashkin70,ashkin86} that won Nobel Prizes
for Steven Chu in 1997 and Arthur Ashkin in 2018.

\section{The Importance of Plasma $\beta < 1$}
\begin{figure}[ht!]
\epsscale{1.2}
\plotone{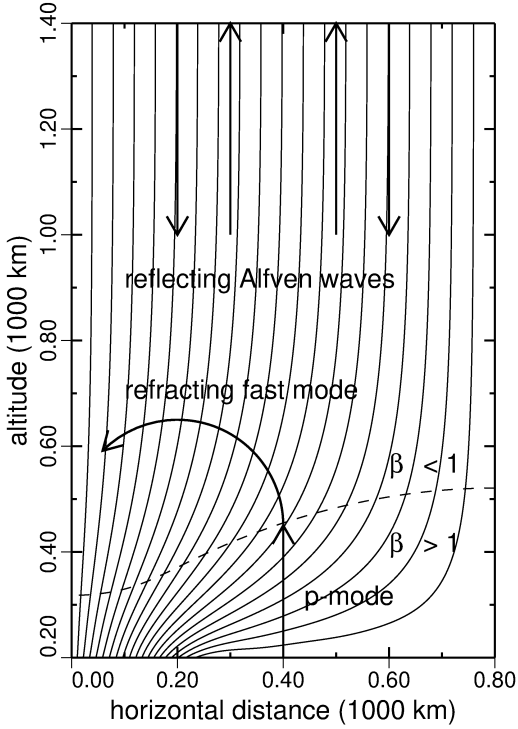}
\caption{Schematic diagram of chromospheric magnetic field lines (solid), equipartition layer (dashed) dividing the 
$\beta = 8\pi nk_{\rm B}T/B^2 {> \atop  <} 1$ regions, and wave propagations giving rise to 
competing FIP and Inverse FIP fractionations.}
\label{fig:beta}
\end{figure}

Figure \ref{fig:beta} shows a schematic diagram of magnetic fields \citep[calculated from][]{athay81} and wave propagation in the solar 
chromosphere, indicating the key components. Upcoming p-modes mode convert at the layer where sound and Alfv\'en speeds are equal 
($\beta = 6/5$ in $\gamma =5/3$ gas) and propagate into the $\beta < 1$ region as fast mode waves, where they undergo a total internal reflection.
Alfv\'en waves of coronal origin enter from the top, and are reflected back upwards, and competition between these two wave fields gives the
Inverse FIP or FIP fractionation. Here we give some further justification for why 
fractionation must occur in this region, and not below this layer, even though magnetosonic and Alfv\'en waves may still propagate there. 
The ion-neutral separation caused by the ponderomotive force must compete with mixing caused by turbulence.
Wave-wave interactions throughout the solar atmosphere can cause a cascade to successively 
shorter wavelengths and ultimately to fundamental length scales like the mean free path or the gyroradius that can cause microscopic mixing.
In the solar photosphere hydrodynamic turbulence dominates, with cascade rate $\simeq \delta u/L$ \citep[e.g.][]{batchelor86,landau87} 
where $\delta u$ is the velocity fluctuation amplitude and $L$ is the eddy scale length. Taking $\delta u\simeq 0.1$ km s$^{-1}$ \citep{reardon08} and 
$L=2000$ km s$^{-1}$ to be the wavelength of a 5 minute sound wave gives a cascade rate of $5\times 10^{-5}$ s$^{-1}$.

With ponderomotive acceleration $a$, the fractionation is given by \citep[e.g.][]{laming15}
\begin{eqnarray}
\nonumber f_k&=&{\rho _k\left(z_u\right)\over\rho _k\left(z_l\right)}\\ &=&\exp\left\{
\int _{z_l}^{z_u}{2\xi _ka\nu _{kn}/\left[\xi _k\nu
_{kn} +\left(1-\xi _k\right)\nu _{ki}\right]\over 2k_{\rm B}T/m_k+v_{||,osc}^2+2u_k^2}dz\right\},
\end{eqnarray}
derived from the momentum equations for ions and neutrals in
a background of protons and neutral hydrogen for densities of element $k$ above ($\rho\left(z_u\right)$)
and below ($\rho\left(z_l\right)$) the fractionation region. In these equations $\xi _k$ is the 
ionization fraction of element $k$, $\nu _{ki}$ and $\nu _{kn}$ are collision frequencies of
ions and neutrals of element $k$ with the background gas \citep[mainly hydrogen and protons,
given by formulae in][]{laming04}, $k_{\rm B}T/m_k \left(
=v_z^2\right)$ represents the square of the element $k$ thermal velocity along
the $z$-direction, $u_k=\partial z/\partial t$ is the upward flow speed and $v_{||,osc}$ a
longitudinal oscillatory speed, corresponding to upward and downward propagating sound waves.
The fractionation rate is then given by
\begin{equation}
{\partial\ln f_k\over\partial t}\simeq {2\xi _ka\nu _{kn}/\left[\xi _k\nu
_{kn} +\left(1-\xi _k\right)\nu _{ki}\right]\over 2k_{\rm B}T/m_k+v_{||,osc}^2+2u_k^2}u_{k}.
\end{equation}

For closed loops supporting resonant Alfv\'en waves, the ponderomotive acceleration is strongest in the strong density gradient at the 
top of the chromosphere with value $a\simeq 10$ km s$^{-2}$. Then with $\xi _k\simeq 1$, $u_k\simeq 0.1$~km s$^{-1}$, and 
$v_{||,osc}\simeq 10$ km s$^{-1}$ the fractionation rate is of order 0.01 s$^{-1}$. Lower down in the chromosphere, closer to the
$\beta =1$ layer, $a\simeq 0.1$ km s$^{-2}$ and $u_k\simeq 0.001$ km s$^{-1}$ giving a rate of order $10^{-6}$~s$^{-1}$,
so clearly such fractionation would not survive against photospheric turbulence.

Alfv\'en and fast mode waves in the $\beta < 1$ chromospheric regions may also generate turbulence and mixing, though compared to the
hydrodynamic case, the cascade rate is lower \citep[e.g.][]{ng97} because the interaction time, $1/k_{||}v_A$, is usually much less 
than the eddy turn over time, $1/k_{\perp}\delta v_{\perp}$, where the subscripts $||$ and $\perp$ refer to the velocity fluctuations and wave vectors parallel to and perpendicular to the magnetic field. If this then requires $N=\left(k_{||}v_A/k_{\perp}\delta v_{\perp}\right)^2$
interactions to complete the eddy turn over, 
the cascade rate in weak turbulence is given by the heuristic expression $k_{||}v_A/N = \left(k_{\perp}\delta v_{\perp}\right)^2/k_{||}v_A$, \citep[e.g.][]{beresnyak19}. A derivation of this expression from quasi-linear plasma wave theory is sketched in the Appendix. 
The expressions derived in the appendix (Equation A10) and that quoted above
coincide for $k_{\perp}/k_{||}=1/4\sqrt{2}$. At the top of the chromosphere where the fractionation
rate is maximized, $\delta v_{\perp}\sim 10$ km s$^{-1}$, $v_A\sim 100$ km s$^{-1}$ and the cascade rate for 5 minute waves is $1/32\times 0.02\times 10^2/100^2 \sim 10^{-5}$ s$^{-1}$, to be compared with a fractionation rate of 0.01 s$^{-1}$ estimated above. 
Closer to the $\beta =1$ layer $\delta v_{\perp}\sim 0.1$ km s$^{-1}$, $ v_A\sim 6$ km s$^{-1}$, and the cascade rate evaluates to 
$1/32\times 0.02\times 0.1^2/6^2 \sim 10^{-7} - 10^{-6}$ s$^{-1}$, comparable to fractionation rate.

\section{The Origin of the FIP and Inverse FIP Effects: All Models are Wrong, but Some are Useful!}
\label{sec:origin}
Wave-wave interactions and turbulence are the most commonly discussed phenomena by which waves modify plasma behavior. The ponderomotive
force is seemingly less well known. 
The ponderomotive force due to optical light was originally invoked by Kepler following the return in 1607 of what we now know to have 
been Halley's comet
to (incorrectly) explain the direction of the comet tail in the solar radiation field \citep{ridpath85}\footnote{John Herschel observing the 1835 return of Halley's comet provided the correct explanation, the solar wind!}.  
The ponderomotive force associated with Alfv\'en waves has mainly seen applications in auroral and magnetospheric physics 
\citep[see e.g. the review by][]{lundin06} who also briefly discuss astrophysical examples such as the collimation of jets, and has
been derived in a number of ways. \citet{lundin06} give a derivation considering the single particle motions in the magnetic and electric fields of an 
Alfv\'en wave, where the force arises as the $\delta v_x$ component of the induced particle motion is crossed with the $\delta B_y$ of the wave motion to 
give a force along the ambient magnetic field direction, $\hat{\bf z}$. \citet{laming09,laming15} consider the Lagrangian for plasma particles interacting with an Alfv\'en wave, which requires knowledge of the energy partitioning within the wave. \citet{laming17} derives the ponderomotive force from considerations of the polarization and magnetization induced in a medium by waves,  and \citet{washimi76} work from the change in relative 
permittivity  induced by the passage
of a wave. These last two only require knowledge of the wave dielectric constant. The derivations are collected
in the appendix of \citet{laming17}, and all agree with equation 30 from the more detailed and general mathematical 
treatment given by \citet{lee83}. The time independent part of their expression is reproduced here,
\begin{eqnarray}
\nonumber f_j&=&{1\over 16\pi}\bigg[\left(\epsilon _{\beta\alpha}-\delta _{\beta\alpha}\right){\partial\delta E_{\alpha}\delta E^*_{\beta}\over\partial x_j}\\
\nonumber &+&{\partial\over\partial x_l}\left(\varepsilon _{jmp}\varepsilon _{klm}B_p{\partial\epsilon _{\beta\alpha}\over\partial B _k}\delta E_{\alpha}\delta E^*_{\beta}\right)\bigg]\\
\nonumber &=& {1\over 16\pi}\left[\left(\epsilon _{\beta\alpha}-\delta _{\beta\alpha}\right){\partial\delta E_{\alpha}\delta E^*_{\beta}\over\partial x_j}
+{\partial\epsilon _{\alpha\beta}\over\partial {\bf B}}{\partial {\bf B}\over\partial x_j}\delta E_{\alpha}\delta E^*_{\beta}\right]\\
&=&{\rho c^2\over 2}{\partial\over \partial x_j}\left(\delta E_{\perp}\delta E^*_{\perp}\over B^2\right),
\end{eqnarray}
where $\epsilon _{xx} = \epsilon _{yy} =  \epsilon _{\perp\perp} = 1+4\pi\rho c^2/B^2$ is the plasma relative permittivity for $\delta {\bf E}$ perpendicular to the background magnetic field ${\bf B} = B\hat{\bf z}$, 
$\delta _{\beta\alpha}$ is the Kronecker delta, and $\varepsilon _{jmp}$ is the Levi-Civita epsilon. Summing is implied over repeated indices.

None of these derivations make any reference to the polarization of the wave, and so we expect the same expression
for the ponderomotive force should apply to Alfv\'en and fast mode waves equally (in conditions where the plasma $\beta < 1$), with any differences in fractionation
by these different waves arising from their different propagations through and interactions with other waves in the chromosphere. To emphasize and explore this point, we give here a mathematically less rigorous derivation, 
but with stronger physical motivation following \citet{ashkin70,ashkin86}.

\begin{figure*}[ht!]
\plottwo{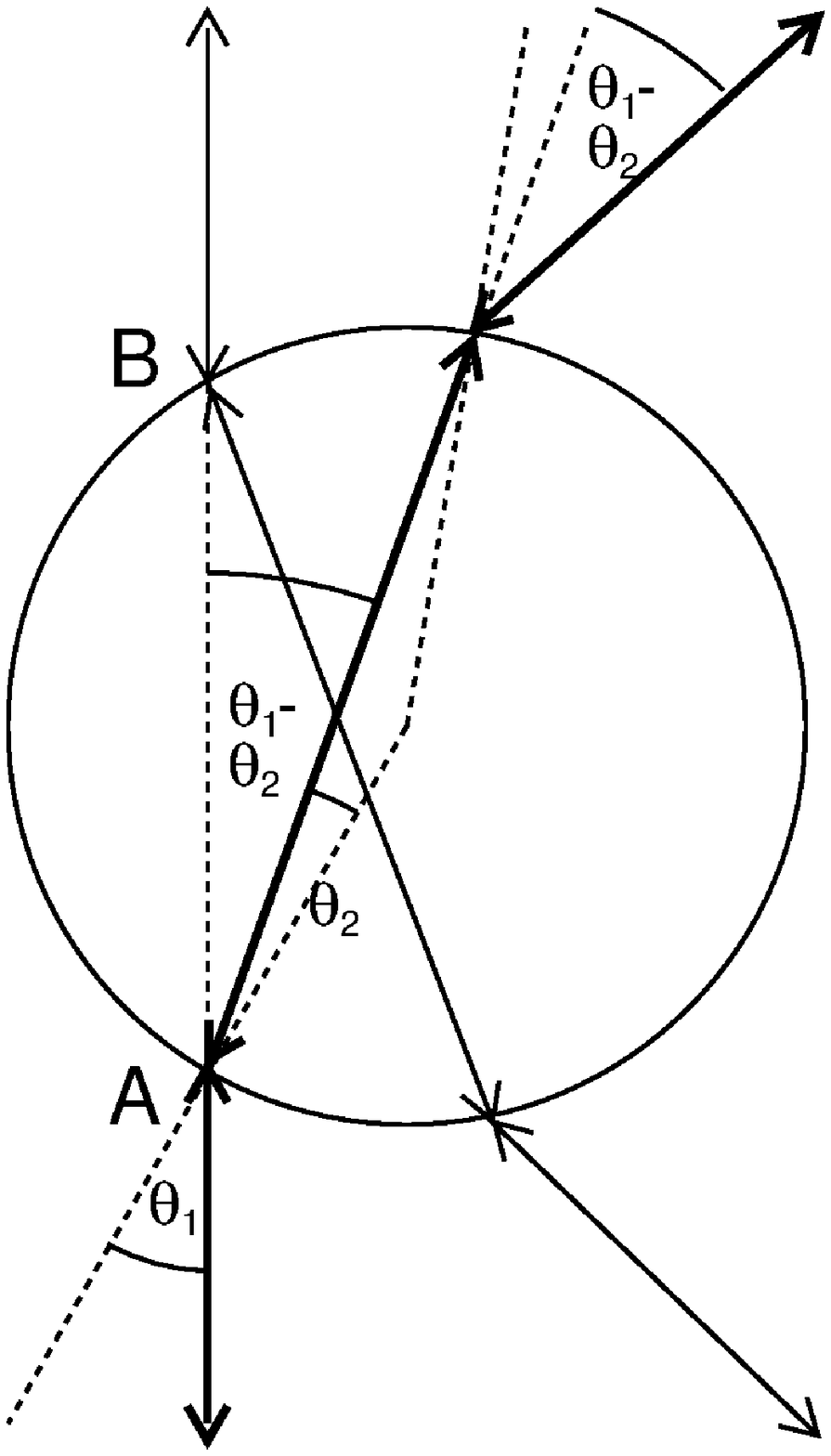}{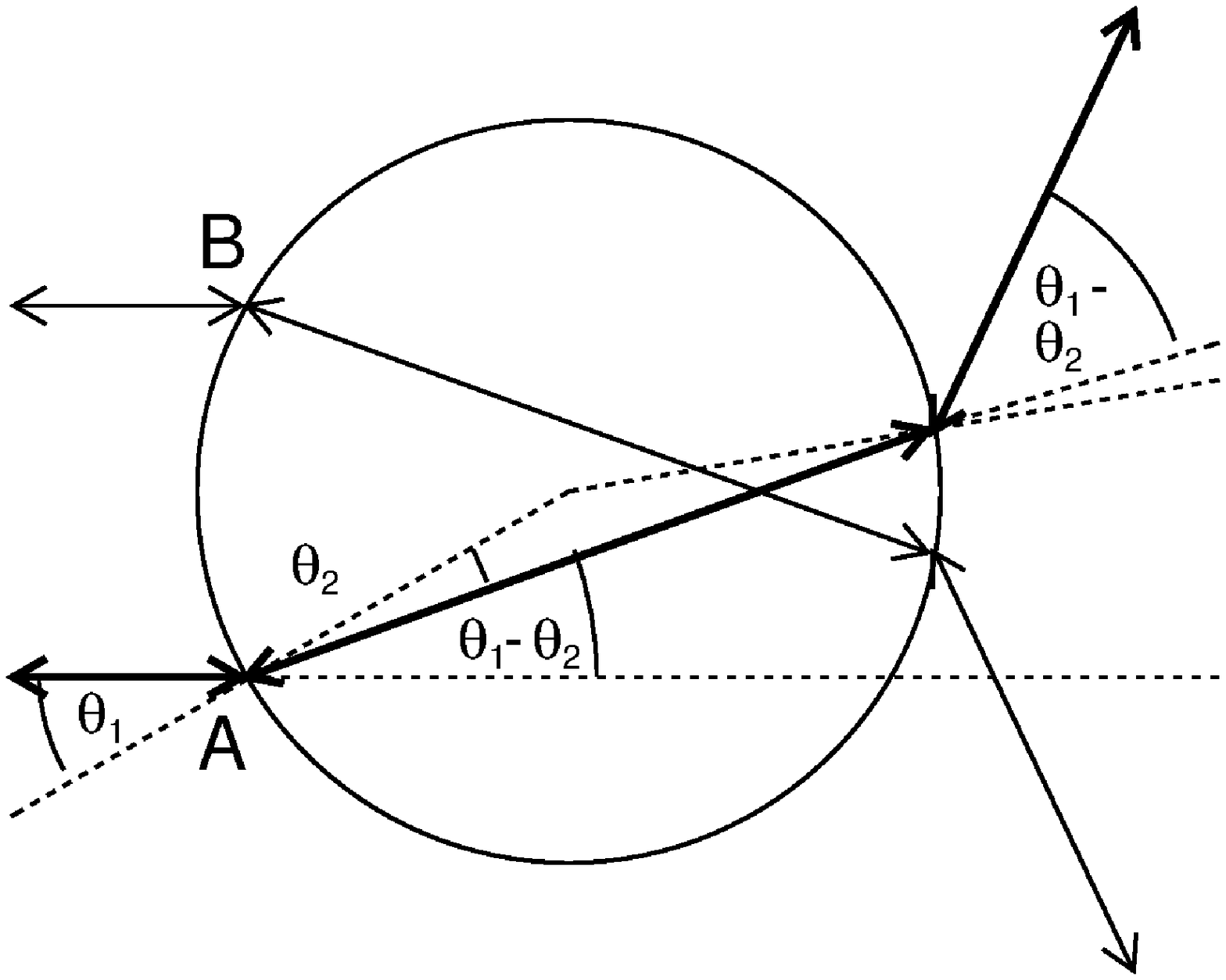}
\caption{Wave motions around a dielectric sphere with refractive index $n_2$, embedded in a medium with refractive index $n_1$. Magnetic field
and wave energy density gradient are in the vertical direction. The left panel shows the effect of parallel propagating waves. The bold arrows show
a wave entering (or exiting) at point A, and being refracted downwards at both entrance and exit. A less energetic wave entering or exiting at point B is 
refracted upwards, but by a lesser amount, so that the net refraction of the wave is still downwards, resulting in an upwards force on the sphere, 
in the direction of $-\partial U/\partial z$, analogous to the FIP Effect. 
The right panel illustrates the case of a perpendicularly propagating fast mode wave. The wave entering at
A is refracted upwards more than the wave entering at B, resulting in a downwards force on the sphere, in the direction of $\partial U/\partial z$. This is
the origin of the Inverse FIP Effect.}
\label{fig:ashkin}
\end{figure*}

Consider first the refraction of fast mode waves around a sphere of higher density and therefore lower Alfv\'en speed embedded in a medium with magnetic
field and gradient of wave energy density in the $\hat{z}$ direction, as in the left panel of Figure \ref{fig:ashkin}. We ignore wave reflection at the boundary 
of the sphere, assume geometrical optics for the wave propagation, and take $\beta << 1$ so that fast modes are essentially isotropic
in their propagation. The initial wave direction makes an angle
$\theta _1$ with the normal to the boundary upon entering the sphere at point A, and is refracted to an angle $\theta _2$, illustrated by the bold ray. 
The total deflection of the ray after refracting at entrance and exit
is $2\left(\theta _1-\theta _2\right)$, for a total longitudinal momentum transfer to the sphere of $\Delta p_{||}=2p\sin ^2\left(\theta _1-\theta _2\right)$,
where $p$ is the initial wave momentum \citep[more properly the wave pseudomomentum,][]{peierls91}. In conditions of a negative wave energy
gradient going upward, a wave with smaller momentum enters at point B, undergoes the same refractions, and would move the sphere downward by a smaller amount, leaving a net upward force.

Writing the spatially varying component of the wave momentum as $p=-r\cos\theta _1 \times \partial U/\partial z/v_A$ where $U$ is the wave energy density and
$v_A$ is the Alfv\'en speed, the rate of momentum change is
\begin{eqnarray}
\nonumber\Delta \dot{p}_{||}&=&-\int _0^{\pi}d\phi\int _{-\pi}^{\pi}2{\partial U\over\partial z}r\cos^2\theta _1\times\\
& & \left(\sin\theta _1\cos\theta _2-\sin\theta _2\cos\theta _1
\right)^2r^2d\theta _1
\end{eqnarray}
which is multiplied out using $n_2\sin\theta _2/n_1\sin\theta _1= 1$ (Snell's Law) to
\begin{eqnarray}
&\Delta &\dot{p}_{||}=-2\pi r^3{\partial U\over\partial z}\int _{-\pi}^{\pi}\sin^2\theta _1-\sin ^4\theta _1+ {n_1^2\over n_2^2}(\sin^2\theta _1\\
\nonumber & &-3\sin^4\theta _1 +2\sin ^6\theta _1)
-2{n_1\over n_2}\sin^2\theta _1\cos^3\theta _1\cos\theta _2d\theta _1.
\end{eqnarray}
In the last term we put $\cos\theta _2\approx\cos\theta _1$ and integrate to find
\begin{equation}
\Delta\dot{p}_{||}=-4\pi ^2r^3{\partial U\over\partial z}\left({1\over 8} - {1\over 8}{n_1\over n_2}\right).
\end{equation}

The calculation for Alfv\'en waves proceeds similarly, except that Snell's Law is now
\begin{equation}
{n_2\sin\theta _2\over n_1\sin\theta _1}=\cos\theta _1\cos\theta _2\ +{n_2^2\over n_1^2}\sin\theta _1\sin\theta _2,
\end{equation}
derived from equation 11 in \citet{laming15b}. This applies for initial waves propagating along the magnetic field. Compared to \citet{laming15b}, 
the magnetic field does not change direction between the two media, so the notation $\left\{\theta _{2w}, \theta _2\right\}$ in \citet{laming15b}
goes over to $\left\{\theta _2, \theta _1\right\}$ here. The second refraction on exit is much more complicated, because the initial wave is no longer
propagating along the magnetic field. We simply approximate this second refraction as one half of the entrance refraction, which then recovers
equation 6 upon integration over angles as before.

For Alfv\'en waves, relating ${\partial U/\partial z}$ to $\delta E$ is relatively straightforward. Since Alfv\'en waves carry their energy along the magnetic 
field, and do not refract in increasing Alfv\'en speed like fast modes,  we take the Poynting vector $S=Uv_A$ for upward and downward going waves
separately and assume $S/B$ to be constant. Then
\begin{equation}
{\partial\over\partial z} \left(S\over B\right) = {\partial U\over\partial z}{v_A\over B}+U{\partial \over\partial z}\left(v_A\over B\right)=0
\end{equation}
so that
\begin{eqnarray}
\nonumber {\partial U\over\partial z}&=&-{UB\over v_A}{\partial \over\partial z}\left(v_A\over B\right)
= -{SB\over v_A^2}{\partial \over\partial z}\left(\delta E^2\over B^2\right){Bc^2\over 8\pi S}\\ \nonumber &&\\
&=& -{\rho c^2\over 2}{\partial \over\partial z}\left(\delta E^2\over B^2\right).
\end{eqnarray}
In the limit $n_2 >> n_1$, with $\rho$ interpreted as the density inside the sphere, the acceleration is given by 
\begin{equation}
a={3\pi c^2\over 16}{\partial \over\partial z}\left(\delta E^2\over B^2\right),
\end{equation}
different from the ``exact'' result in Equation 3 by a factor $3\pi /8$.

In the case of fast mode waves, expressing  $\partial U/\partial z$ in terms of $\delta E^2$ is more involved. We take
\begin{eqnarray}
{\partial I_+\over\partial z}&=&-{I_+\over v_A}{\partial v_A\over\partial z}-I_+{\partial f_R\over\partial z}\\
{\partial I_-\over\partial z}&=&-{I_-\over v_A}{\partial v_A\over\partial z}+I_+{\partial f_R\over\partial z}
\end{eqnarray}
where $I_{\pm}$ are
the intensities of upward and downward going waves and
$f_R=\sqrt{1-2c_s\left(z_{\beta =1}\right)^2/\left(v_A\left(z\right)^2+c_s\left(z\right)^2\right)}$ represents the cumulative fraction of initially
upward going waves refracted back downwards at altitudes $z$ above the layer where $\beta =1$, $z_{\beta =1}$ \citep{wood12,laming15}. Then
forming $\partial\left(I_+-I_-\right)/\partial z$ and $\partial\left(I_++I_-\right)/\partial z$ we find for the cross helicity, $H$, Poynting flux, $S$, and
wave energy density $U$,
\begin{eqnarray}
{\partial H\over\partial z}&=&{\partial \over\partial z}\left(I_+-I_-\over I_++I_-\right)=-{2I_+\over U}{\partial f_R\over\partial z}\\
{\partial S\over\partial z}&=&{\partial \over\partial z}\left(I_+-I_-\right)v_A=-{2I_+v_A}{\partial f_R\over\partial z}\\
{\partial U\over\partial z}&=&{\partial \over\partial z}\left(I_++I_-\right)=-{U\over v_A}{\partial v_A\over\partial z}.
\end{eqnarray}
With $v_A=\left(c^2\delta E^2/8\pi\right)H/S$ and assuming $v_A >> c_s$
\begin{equation}
{\partial\ln H\over \partial z}={\partial\ln S\over\partial z} = {\partial\ln\left(c_s^2/v_A^2\right)\over\partial z},
\end{equation}after putting $I_+=I_0\left(1-f_R\right)$ and $I_-=I_+f_R$. We find
\begin{equation}
{\partial U\over\partial z}=-{c^2\over v_A^2}{\partial\over\partial z}\left(\delta E^2\over 8\pi\right)-{c^2\over c_s^2}{\delta E^2\over 8\pi}
{\partial\over\partial z}\left(c_s^2\over v_A^2\right)+{c^2\over v_A^2}{\delta E^2\over 8\pi }{\partial\ln S\over\partial z}
\end{equation}
where $\gamma P$ in $c_s^2=\gamma P/\rho$ has been assumed constant in the term including $\partial\ln H/\partial z$. Pursuing these 
approximations, this term should cancel with the term in $\partial\ln S/\partial z$ to leave only the first term. Neglecting the change in $S$, 
as we did for Alfv\'en waves, leaves only the first two terms contributing to the ponderomotive force which give Equation 9. Then from Equation 6 in the limit $n_2 >> n_1$ we find Equation 10 once more. Although the approximations in this approach are
somewhat loose, we emphasize that this is the correct expression for the ponderomotive force due to fast modes, to within a factor
$3\pi /8$,  as evaluated with fuller
mathematical rigor from Equation 3, which is the expression used in the modeling below.

Depending on the sign of ${\partial\left(\delta E^2/B^2\right)/\partial z}$ Equations 3 or 10 admit FIP or Inverse FIP effect. 
Since $B$ is usually decreasing
with height, FIP effect (positive gradient) is more common, but if $\delta E^2/B^2$ decreases with height, Inverse FIP fractionation can result.
Such a solution requires strong reflection of initially upward propagating waves back downwards, and is generally only realistic for fast mode waves
undergoing a total internal reflection. The best way to understand the physical basis for this trapping is to consider the now perpendicularly propagating
fast mode waves, as in the right hand panel of Figure 1. The wave entering or exiting at point A refracts more energy upwards than the wave entering or exiting at point B, and hence the sphere is pushed downwards. 
So in a negative wave energy density gradient as before, the sphere is pushed {\em up} the
gradient by the refraction of initially perpendicularly propagating waves, rather than down the gradient by initially parallel propagating waves. 
Calculating the refraction as before yields the negative of equation 6 for $\Delta\dot{p}_{||}$. With zero Poynting vector, we
write $U=\rho c^2\delta E^2/2B^2$ and remember that $\rho$ refers to the density of the sphere (and not the surroundings), so that 
$\partial U/\partial z = \rho c^2/2\times \partial\left(\delta E^2/B^2\right)/\partial z$ which recovers Equation 10. These perpendicularly 
propagating fast modes most likely derive from the total internal reflection of an initially upward propagating wave field.

\section{FIP and Inverse-FIP Modeling for Solar Flares}
\subsection{Preamble}
We return to the observations of solar flares \citep{dennis15,katsuda20} and attempt to understand the abundance patterns seen in terms
of the FIP and Inverse FIP effects on solar loops. We model shear Alfv\'en waves on a 100,000 km loop with 300 G magnetic field. The main 
difference in the models is the degree to which the magnetic field lines expand, and the magnetic field strength decreases, between the photosphere
and the corona. The equipartition layer, where gas and magnetic pressures are equal, lies in the middle of this magnetic field expansion. The
models cited later on are given in Table 1. For constant coronal magnetic field, higher photospheric field results from greater expansion, and the loop resonant frequency is slightly higher due to the reduced travel time at the Alfv\'en speed from one footpoint to the other, although the
equipartition layer lies a little lower. For all models, the Alfv\'en wave
amplitude corresponds to a velocity perturbation of about 130 km s$^{-1}$ in the corona at a density of about $4\times 10^8$ cm$^{-3}$. 
Higher coronal density, $n_e$, would lead to a lower coronal velocity perturbation, proportional to $1/\sqrt{n_e}$ for a standing wave. A uniform
upgoing slow mode wave energy flux of $10^8$ ergs cm$^{-2}$ s$^{-1}$ deriving from photospheric convective motions is injected at the $\beta =1$
layer. The amplitude increases as the density decreases, according to the WKB approximation, until the slow mode wave amplitude reaches the local sound speed. At this point further WKB amplitude growth ceases, on the assumption that a shock will form and radiate away the excess kinetic energy.

\begin{table}
\begin{center}
\caption{Loop Models}
\begin{tabular}{lccC}
\hline
$B_{cor}/B_{phot}$ & ang. freq.  (rad s$^{-1}$) & $h_{\beta =1}$ (km) \\\hline
0.3 & 1.0845& 300 \\
0.5 & 1.073& 350 \\
0.7 & 1.064& 350 \\

\hline
\end{tabular}
\end{center}
\end{table}

\subsection{Fractionation Mechanism}
Our discussion above suggests that the ``usual'' FIP effect should result from the ponderomotive force developed by parallel propagating
Alfv\'en or fast mode waves. These are treated by integrating the Alfv\'en wave transport equations as in previous papers \citep{laming09,laming12,
laming15,laming17,laming19}. The refraction of perpendicularly propagating fast mode waves gives the Inverse FIP effect. These are treated following
\citet{wood12,laming15} as approximately isotropic in the upward moving hemisphere.
At chromospheric height $z$  the reflected fraction is approximately
\begin{equation}
f_R\left(z\right)\simeq\sqrt{1-{2c_S^2\left(z_{\beta =1}\right)\over v_A^2\left(z\right)+c_S^2\left(z\right)}}
\end{equation}
where $z_{\beta =1}$ is the chromospheric height where mode conversion occurs. This is slightly different to the previous versions
cited above in including the factor of 2 multiplying $c_S^2\left(z_{\beta =1}\right)$ in the numerator inside the square root. This gives 
$f_R\left(z_{\beta =1}\right) = 0$ instead of $1/\sqrt{2}$, which reflect a presumed origin of fast mode waves as coming from waves excited by
sub-photospheric reconnection, rather than by convective motions. Convective motions would be ubiquitous, and upgoing and downgoing waves
should be present in any chromospheric region. Waves from reconnection are spatially localized, and do not in general return to the $\beta = 1$ 
layer at the same place where they left it.

We take the ponderomotive acceleration due to fast mode waves to be of the same form as that due to Alfv\'en waves,
\begin{equation}
a={c^2\over 4}{\partial\over\partial z}\left(\delta E^2\over B^2\right)={\delta v^2\over 2}\left(1-f_R\right){1\over\delta v}{\partial\delta v\over\partial z}-{\delta v^2\over 4}
{\partial f_R\over\partial z}
\end{equation}
with variations in the acceleration arising due to the specific wave fields and propagation of the different modes.
The two terms represent an upwards contribution arising as the fast mode waves increase
in amplitude as they propagate through progressively lower density plasma, and the downwards contribution
arising from fast mode wave reflection. Evaluating
\begin{eqnarray}\nonumber {\partial f_R\over\partial z} &=&{c_S^2\left(z_{\beta =1}\right)v_A\over\left(v_A^2+c_S^2\right)^2f_R}{\partial v_A\over\partial z}\\
&=& {c_S^2\left(z_{\beta =1}\right)v_A^2\over\left(v_A^2+c_S^2\right)^2f_R} \left({1\over H_B}-{1\over 2H_D}\right)
\end{eqnarray}
and assuming from the WKB approximation
\begin{eqnarray}
{1\over \delta v}{\partial \delta v\over\partial z}&=&{-1\over 2H_B}-{1\over 4H_D},\\
\nonumber
\end{eqnarray}
where $H_D$ and $H_B$ are the signed density and magnetic field scale heights, we find from Equation 19
\begin{eqnarray}
\nonumber a&=&{\delta v^2\over f_R}\bigg\{\left(f_R-1\right)\left(-{1\over 8H_D}-{1\over 4H_B}\right)\\
&+&{c_S^2\left(z_{\beta =1}\right)\over\left( v_A^2+c_S^2\right)^2}\left(-{c_S^2\over 8H_D} -{c_S^2\over 4H_B}-{v_A^2\over 2H_B}\right)\bigg\}.
\label{eq:fastmodepond}
\end{eqnarray}
With both $H_D$ and $H_B$ negative and $f_R < 1$, the first term in curly
brackets is negative, giving Inverse FIP effect. The second term is positive, giving positive FIP effect.
In conditions where $v_A \gg c_S$, an overall downwards pointed ponderomotive acceleration requires 
$\left|H_D\right| < \left|H_B\right|/2$ in this simple model.

The reflection of fast mode waves from other perturbations, e.g., density fluctuations caused by other waves or shocks (not included in this model) would increase the Inverse FIP effect developed. Even so, Equation \ref{eq:fastmodepond} suggests that Inverse FIP effect is more likely to found
in conditions with minimal magnetic field expansion through the chromosphere. This might correspond with the conclusions of \citet{baker19,baker20}
that sub-photospheric reconnection is required to generate the necessary waves to cause Inverse FIP, thus weakening the magnetic field low down 
in the atmosphere relative to that higher up. It might also fit with
observations in M~dwarfs. While the magnetic fields measured in these stars are similar to those
in the Sun, the filling factor is higher \citep[e.g.,][]{donati09,reiners09}, allowing less volume for
expansion with increasing altitude.

\begin{table*}
\begin{center}
\caption{Data and Models for \citet{katsuda20}}
\begin{tabular}{lc|cccc|cccc|cccc}
\hline
 &  & \multicolumn{4}{c}{$B_{cor}/B_{photo} =0.3$}& \multicolumn{4}{c}{0.5} & \multicolumn{4}{c}{0.7}\\
ratio & obs. abundance &$v_{fm}=0$& 9& 12& 15& 0& 8& 10& 12& 0& 6& 7& 8\\\hline
Si/H&  0.76 [0.38 - 1.09]&   2.71& 2.13& 1.77& 1.39&  2.55& 1.22& 0.80& 0.48& 2.79& 1.26& 0.95& 0.68 \\  
S/H&   0.39  [0.21 - 0.49]&  1.33& 1.11& 0.96&  0.81&  1.26& 0.67& 0.58& 0.31& 1.35& 0.68& 0.54& 0.40\\  
Ar/H&  0.86 [0.55 - 1.29]&   1.08& 1.08& 1.08&  1.08& 1.03& 1.03& 1.03& 1.03& 1.11& 1.11& 1.11& 1.11\\  
Ca/H&  1.89 [1.61-  2.29]&  5.07& 3.98& 3.30&  2.59& 4.79& 2.07& 1.29& 0.73& 5.32& 2.03& 1.55& 1.06\\   
Fe/H& &                             4.62& 3.64& 3.02& 2.39& 4.37& 1.74& 1.04& 0.55& 4.84& 1.77& 1.23& 0.81\\   

\hline
\end{tabular}
\end{center}
\tablecomments{Fast mode wave amplitudes at $\beta =1$ are in km s$^{-1}$. Coronal magnetic field $B_{cor}=300$ G. Abundance ratios 
are averages of 10 spectra from 4 flares, relative to photospheric abundances of \citep{lodders03}. 
Ranges are minimum and maximum values from DEM analysis.}
\end{table*}

\subsection{Charge Exchange}
In modeling Inverse FIP fractionation, the ionization balance in the low chromosphere, where the ponderomotive force due to the fast mode
waves develops, becomes crucial. This is dominated for many elements by 
charge exchange ionization by protons. We take most rates as the minimum of either the Langevin estimate \citep[see][]{laming12} 
or the rate given by \citet{kingdon96}. The Langevin rate is taken to be a theoretical maximum, so a rate larger than this must be in error, but the rate may be smaller given sufficient scarcity of final states for the captured electron. 
The Langevin estimate is also taken where rates are missing from the
\citet{kingdon96} tabulation. Additionally, we take the charge exchange ionization of S from \citet{zhao05}. 
Prior work \citep[e.g.][and its antecedents]{laming19} had used the rate of
\citet{butler80} and \citet{leteuff00} in preference to the result tabulated by \citet{kingdon96}, the discrepancy being about five orders of magnitude.
At chromospheric temperatures, the rate of \citet{zhao05} is about three orders of magnitude larger than \citet{kingdon96}, and two orders of magnitude
lower than in our previous calculations. The ionization fraction of S is correspondingly lower than in for example \citet{laming19}, but still sufficiently high 
that S is subject to the Inverse FIP effect. We also replace the charge exchange ionization rate for Na with the Langevin estimate, while the corresponding rate for Fe is reduced from the value given by \citet{kingdon96}.
The polarizabilities coming into the charge exchange rates, and also into the elastic scattering rates involving neutrals are taken from the
tabulation of static polarizabilities of \citet{schwerdtfeger19}.

\subsection{Results}
\citet{katsuda20} collect 10 spectra from 4 X-class flares observed in 2005, and derive abundances from each spectrum using fits with two plasma components at different temperatures, and for comparison also employing a full differential emission measure (DEM) treatment. Coronal
abundances are compared with photospheric values from \citet{lodders03}. We take the average of the abundances
deduced by the DEM method, and compare them with various models in Table 2. The Si abundance is taken to be the average of the abundances 
derived from Si XIII He$\alpha$ and Si XIV Lyman $\alpha$, while the S, Ar, and Ca abundance come from the He$\alpha$ transitions of the He-like
ion of each element. The models in Table 1 are given for ratios of the coronal to photospheric magnetic field $B_{cor}/B_{photo} = 0.3$, 0.5, and 0.7.
For each value of $B_{cor}/B_{photo}$, abundances relative to H are tabulated for different values of the upgoing fast mode wave amplitude entering
the $\beta < 0$ region of the chromosphere at the equipartition layer. For zero fast mode wave amplitude, a standard FIP effect results due to the 
Alfv\'en waves in the model. As the fast mode wave amplitude is increased, the degree of FIP fractionation reduces, eventually becoming Inverse FIP
at the highest fast mode wave amplitudes considered, for $B_{cor}/B_{photo} = 0.5$ and 0.7. The best match is probably achieved for
$B_{cor}/B_{photo} =0.7$ with 7 km s$^{-1}$ fast mode wave amplitude. S exhibits the strongest Inverse FIP effect, both in observations and
in models. Si also shows Inverse FIP
but to a lesser degree than S, Ar is essentially unchanged, while Ca retains positive FIP effect. For reference, and for comparison with
the results of \citet{dennis15} to be discussed below, we also give the fractionations for Fe/H.

\begin{figure*}[ht!]
\epsscale{1.2}
\plotone{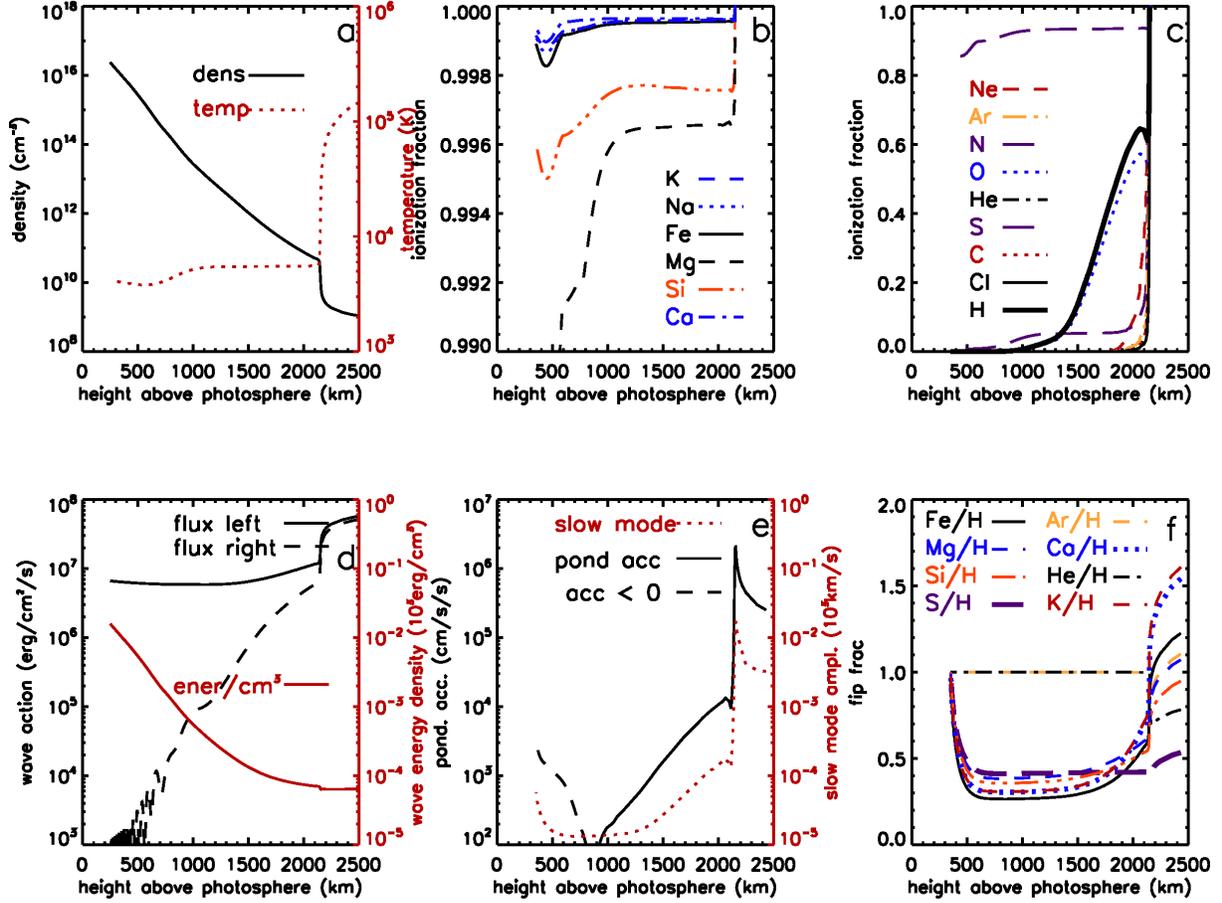}
\caption{Chromospheric model and wave fields for FIP - Inverse FIP fractionation, shown for magnetic field expansion $B_{cor}/B_{photo} =0.7$, fast mode wave speed at 350 km ($\beta =1$) 7 km s$^{-1}$. Panel a gives density and temperature profiles, b gives ionization fractionation for low FIP elements,
c ionization fractions for high FIP elements, d gives Alfv\'en wave energy fluxes for downgoing (left) and upgoing (right) waves in black, total wave 
energy in red, e gives total ponderomotive acceleration from both Alf\'ven (positive contribution) and fast mode waves (negative contribution) in black, parametrically driven slow mode wave amplitude in red, panel f gives element fractionations.}
\label{fig:frac}
\end{figure*}

To illustrate more clearly the processes in our models, we show in Figure \ref{fig:frac} the various features of the chromospheric model and wave fields that
give rise to these fractionations. Top left (panel a) shows the background chromospheric model density and temperature profiles from \citet{avrett08}.
Top middle (panel b) and top right (panel c) show the ionization fractions for various low FIP and high FIP elements respectively. The change in the
S charge exchange ionization rate has reduced the S ionization fraction from above 99\% to about 94\% in the upper regions of the chromosphere, falling 
to about 85-90\% near the equipartition layer. There are of course several other processes that contribute to the S ionization balance, so that a two order of magnitude change in the charge exchange ionization rate results in only a $\sim 10$\% change in the ionization fraction. 

Bottom left (panel d) shows the
upward and downward Alfv\'en wave energy fluxes (in black), and the Alfv\'en wave energy density in red. Even though the Alfv\'en waves are chosen
to be resonant with the loop, and have the largest amplitude in the coronal section of the model (not shown), the wave energy density is highest in the
chromosphere and is monotonically declining with height. According to our discussion above in section 3, these are conditions where parallel propagating
waves should give positive FIP effect. The strongest negative wave energy gradient is found in a short region at about 2150 km altitude, corresponding
to the steep density and temperature gradients in the background chromospheric model in panel a. Bottom middle (panel e) shows the sum of the 
ponderomotive accelerations (in black), due to parallel propagating Alfv\'en waves giving a positive acceleration, and perpendicularly propagating fast
mode waves resulting form the total internal reflection, giving negative acceleration, leading to Inverse FIP. The slow mode wave amplitude resulting
from parametric excitation by the Alfv\'en wave driver is shown by the red dotted line. This appears squared in the denominator of the integrand 
in equation 1. Finally at bottom right (panel f) we show the resulting fractionations. All the low FIP elements and S are depleted just above the
equipartition layer at 350 km by about the same amount by the negative ponderomotive acceleration associated with the fast mode wave total internal reflection. The other high FIP elements are not affected, being almost completely neutral at this altitude. At the top of the chromosphere, positive FIP
effect acts on different ions in different amounts, depending sensitively on their ionization fractions, leading to overall positive FIP effect for Ca, and Inverse FIP for Si and S.

These effects can be understood with reference to Equation 1. In neutral background gas, the collision frequencies of ions and neutrals with the background,
$\nu _{kn}\sim\nu _{ki}$, and the integrand depends simply on the ionization fraction $\xi _k$. Thus, the low FIPs and S, which all have $\xi _k > 0.8$ 
are fractionated approximately the same in such conditions, and end up taking on similar Inverse FIP effects in the low chromosphere. Towards the top 
of the chromosphere where H is becoming ionized, $\nu _{kn}\ll\nu _{ki}$. Unless $\xi _k$ is very close to unity, to the order of $10^{-3}$ (approximately the ratio $\nu _{kn}/\nu _{ki}$), the fractionation is suppressed by the back diffusion of neutrals. In such conditions, FIP fractionation favors the most
highly ionized elements. Correspondingly, Ca is fractionated the most, followed by Fe and Si. S, ionized to about 95\%, is not FIP fractionated at all
in this region.

\begin{table*}
\begin{center}
\caption{Data and Models for \citet{dennis15}}
\begin{tabular}{lcccccc}
\hline
ratio &  \multicolumn{3}{c}{observations}&  \multicolumn{3}{c}{models for $B_{cor}/B_{photo}=0.7$}\\
&  RESIK & SAX& SAX& 160 km s$^{-1}$& 193 km s$^{-1}$& 225 km s$^{-1}$\\
&  RHESSI& 2007 June 1& 2007-13& 7 km s$^{-1}$& 9 km s$^{-1}$& 10 km s$^{-1}$\\\hline 
Si/H&  2.34& 1.09& 1.64$\pm 0.66$& 1.46& 1.15& 1.02\\
S/H&   1.10& 0.78& 1.23$\pm 0.45$& 0.58& 0.35& 0.23\\
Ar/H&  1.10& 1.48& 2.48$\pm 0.90$& 1.13& 1.17& 1.00\\
K/H&   6.76& & & 3.59& 3.77& 4.74\\
Ca/H&  3.89& 3.59& 3.89$\pm 0.76$& 3.39& 3.48& 4.30\\ 
Fe/H&  2.57& 1.70& 1.66$\pm 0.34$& 2.56& 2.28& 2.50\\

\hline
\end{tabular}
\end{center}
\tablecomments{Fast mode wave amplitudes at $\beta =1$ are in km s$^{-1}$. Abundance ratios are averages
of 526 flares, relative to photospheric abundances of \citep{asplund09}.}
\end{table*}

\citet{dennis15} measure abundances for Si, S, Ar, Ca and Fe from 526 flares observed by the Solar Assembly for X-rays (SAX), part of the X-ray Spectrometer \citep[XRS:][]{schlemm07} on the MESSENGER mission. Table 3 gives their results for the 2007 June 1 flare, for the
average of 526 flares observed between 2007 and 2013, and results for coronal abundances of Si, S, Ar, and K from the RESIK X-ray crystal spectrometer flown on the {\it Coronas-F} satellite \citep{sylwester14} and for Fe from {\it RHESSI} observations \citep{phillips12}. These
coronal abundances are compared with photospheric values from \citet{asplund09}. These are in general lower than the results of
\citet{lodders03}, \citep[used by][]{katsuda20}, by amounts ranging from 5\% for Ca/H to 30\% for Ar/H, leading to higher fractionations reported
by \citet{dennis15}, relative to \citet{katsuda20}, by amounts comparable to or less than the reported fractionation uncertainties.
\citet{dennis15} speculate
that the division between low FIP and high FIP elements may move in flares from the usual 10 eV (i.e. those elements that can be ionized by
H I Lyman $\alpha$) to a value more like 7 eV, but do not specify a mechanism. We argue instead that the peculiar abundance enhancement pattern
is a combination of Inverse FIP effect low down in the chromosphere, a feature only present in solar flares, and the more usual FIP effect at the 
top of the chromosphere where the background gas is becoming ionized. Table 3 compares the abundance fractionations from \citet{dennis15} with
models calculated for  $B_{cor}/B_{photo}=0.7$, with varying coronal Alfv\'en wave amplitude (160 - 225 km s$^{-1}$) and fast mode wave
amplitude at the equipartition layer (7 - 10 km s$^{-1}$). \citet{dennis15} only see an Inverse FIP for S during the 2007 June 1 flare, but in all other
cases S is the least enhanced. But combined with the insights gained from the analysis of \citet{katsuda20}, the combination of FIP and Inverse FIP 
offers a plausible explanation of these observations. Similarly, \citet{warren14} measured an Fe abundance enhancement of $1.17\pm 0.22$ from
640 spectra taken during 21 flares during the solar maximum of 2011-13, using the Solar Dynamics Observatory Extreme-ultraviolet Variability Experiment (SDO/EVE). With just one datum, \citet{warren14} argued that during flares, chromospheric evaporation must originate deep in the chromosphere,
below the region where FIP fractionation occurs. But the fractionation observed is also consistent with our modeling in Table 2 where the \citet{katsuda20}
observations make it clear the Inverse FIP effect is at work.

\section{Discussion and Conclusions}
\label{sec:discussion}
There are many possible processes that might reduce the degree of FIP fractionation from the usual factor of about 3 seen in active region and the
solar wind to lower values seen in solar flares. Either a lower Alfv\'en wave amplitude, or a significant upward flow velocity 
\citep[$u_k$ in Equation 1;][]{laming17} and its associated effects on fractionation and slow mode wave production, could reasonably be expected in solar flare plasmas. However Inverse FIP
fractionation as seen by \citet{doschek15,doschek16} in localized regions and by \citet{katsuda20} in spatially unresolved solar flare plasma requires
a quite different explanation. One of the early attractive features of the ponderomotive force \citep{laming04} was that given suitable wave fields,
it could point upward or downward, giving FIP or Inverse FIP fractionation respectively. Calculations with parallel propagating Alfv\'en waves have
difficulty achieving the degree of wave reflection required to give a plausible Inverse FIP effect. Fine-tuning of wave frequencies far beyond what is
realistic for the solar chromosphere is required, which then leads to the idea of the total internal reflection of fast mode waves as a means of 
generating downward pointing ponderomotive force.

The fast modes most likely originate as sound waves from the photosphere that mode-convert into fast modes at the equipartition
layer. However the fast mode wave amplitudes quoted above at the equipartition layer are significantly larger than those expected to derive
from waves associated with photospheric convection. In our models above, an acoustic wave energy flux of $10^8$ ergs cm$^{-2}$ s$^{-1}$
gives an acoustic wave amplitude of order 1 km s$^{-1}$ at a density of $2\times 10^{16}$ cm$^{-3}$ at an altitude of 300 km, comparable to
amplitudes observed \citep{stangalini12}, whereas fast mode wave amplitudes of several times this value are required in Tables 2 and 3.
\citet{baker19} identified the small patches of Inverse FIP plasma observed in the corona during a flare with sub-photospheric reconnection between 
two interacting sunspots. \citet{baker20} investigate further, and find regions of Inverse FIP effect plasma coinciding with strong light bridges within
sunspot umbrae. \citet{kigure10} have investigated MHD wave generation by reconnection, and following this work, the geometry of reconnection
at the light bridge suggests that mainly Alfv\'en waves, not sound waves, should have been produced. These would propagate upward along the
magnetic field through the equipartition layer without mode conversion here to fast modes. Such mode conversion can happen elsewhere 
if the inclination of the
magnetic field to the vertical direction rotates, such that the Alfv\'en wave polarization is rotated to become fast mode. 

Flare plasma arrives in the corona through the process of chromospheric evaporation. As mentioned above, the term in $u_k$ representing the flow velocity of element $k$ reduces the fractionation when this begins to dominate the
denominator in the integrand of Equation 1, which typically occurs in the range 1 - 10 km s$^{-1}$. Such flow speeds (or higher) are more
likely to be realized in the lower density gas in the upper layers of the chromosphere (assuming constant evaporative mass flux) and so we might 
expect the FIP fractionation to be more affected than the Inverse FIP fractionation, because this last process occurs in the lower chromosphere where flow speeds will be lower. In fact upward flow speeds are likely to be significantly higher than 1 - 10 km s$^{-1}$, and the most plausible scenario for
FIP fractionation seems to be case modeled by \citet{dahlburg16}. Explosive events in the flaring loop release heat and Alfv\'en waves. In conditions
where the Alfv\'en speed is greater than the electron thermal speed, corresponding to minimum magnetic field strengths in the range 30 - 100 G, the 
Alfv\'en waves reach the chromosphere first and fractionate the plasma before the heat flux arrives to cause the evaporation. An episodic heating
process should allow fractionation with large evaporative velocities, and suggest that active regions should show stronger FIP fractionation than
quiet solar corona where the magnetic field is weaker.

The ponderomotive acceleration model of the FIP effect also appears to provide a satisfactory explanation of the Inverse FIP bias. Nothing more exotic
than a source of fast mode waves at the equipartition layer, and the apparent shift of the boundary between low FIP and high FIP elements arises
naturally. Our results indicate that Inverse FIP fractionation in flares could be common, having been generally masked by the competing 
FIP fractionation, possibly suggesting a role for sub-photospheric reconnection in the flare process.

\acknowledgements This work has been supported by the NASA Heliophysics Guest Investigator (80HQTR19T0029) and Supporting Research 
Programs (80HQTR20T0076), and by Basic Research Funds of the Office of Naval Research.

\appendix
\section{Turbulent Cascade Rate from Quasi-Linear Theory}
In anisotropic turbulence, the three wave interaction for $P + Q \longleftrightarrow M$ with wavevectors ${\bf k}_1$, ${\bf k}_2$, and
${\bf k}$ is allowed if $\omega _P\simeq\omega _M$ and $\omega _Q\rightarrow 0$. 
We sketch the calculation of this interaction rate from quasi-linear plasma theory, 
following \citet{melrose86} and \citet{luo06}. The rate equation for
the evoluation of the number of plasma quanta in mode $M$, $N_M$, is
\begin{eqnarray}
{dN_M\left({\bf k}\right)\over dt}&=&4\left(4\pi\right)^3\hbar\int{R_M\left({\bf k}\right)R_P\left({\bf k}_1\right)R_Q\left({\bf k}_2\right)\over
\omega _M\left({\bf k}\right)\omega _P\left({\bf k}_1\right)\omega _Q\left({\bf k}_2\right)}\left|{\bf e}_{Mi}{\bf e}_{Pj}{\bf e}_{Ql}\alpha _{ijl}\right|^2
N_P\left({\bf k}_1\right)N_Q\left({\bf k}_2\right)\left(2\pi\right)^4\\
& & \times \left(2\pi\right)^4\delta ^4\left(k-k_1-k_2\right)
\delta\left(\left|{\bf k}\right| - \left|{\bf k}_1\right|\right){d^3{\bf k_1}\over\left(2\pi\right)^3}{d^3{\bf k_2}\over\left(2\pi\right)^3}
\end{eqnarray}
where $R_M = v_A^2/2c^2$ etc is the ratio of electric to total energy in the wave, and ${\bf e}_{Mi}$ etc are the wave polarization vectors.
The second delta function $\delta\left(\left|{\bf k}\right| - \left|{\bf k}_1\right|\right)$ incorporates the requirement 
$\omega _Q\rightarrow 0$. From \citet{melrose86}
Equation 10.9 where $\omega _Q << \omega _M$, $\omega _P$, we put for the quadratic response tensor
\begin{equation}
\alpha _{ijl}=-{e^3n\over 2m^2\omega _Q}\left\{k_r\tau _{rl}\left(\omega _Q\right)\tau _{ij}\left(\omega _P\right)-
k_{1r}\tau _{rl}\left(\omega _Q\right)\tau _{ij}\left(\omega _M\right)\right\},
\end{equation}
where in the limit $\omega <<\Omega = eB/mc$,  the ion cyclotron frequency,
\begin{equation}
\tau _{ij}\left(\omega\right)=\begin{bmatrix}
-{\omega ^2\over\Omega ^2} & -i\epsilon{\omega\over\Omega} & 0\\
i\epsilon{\omega\over\Omega} &-{\omega ^2\over\Omega ^2} &  0\\
0 & 0 & 1 \end{bmatrix}.
\end{equation}
We choose ${\bf k}=\left(k_{\perp},0,k_{||}\right)$, ${\bf k}_1=\left(k_{1\perp}\cos\phi _1,k_{1\perp}\sin\phi _1,k_{1||}\right)$, and 
${\bf k}_2=\left(k_{2\perp}\cos\phi _2,k_{2\perp}\sin\phi _2,k_{2||}\right)$ so with ${\bf k}= {\bf k}_1+{\bf k}_2$
\begin{equation}
\alpha _{ijl}\simeq -{e^3n\over 2m^2\omega _Q}\left\{k_{2\perp}\cos\phi _2\tau _{xl}\left(\omega _Q\right)\tau _{ij}\left(\omega _P\right)
+k_{2\perp}\sin\phi _2\tau _{yl}\left(\omega _Q\right)\tau _{ij}\left(\omega _M\right)\right\}.
\end{equation}
With  ${\bf e}_M=\left(1,0,0\right)$, ${\bf e}_P=\left(\cos\phi _1,\sin\phi _1,0\right)$, and 
${\bf e}_Q=\left(\cos\phi _2,\sin\phi _2,0\right)$
\begin{equation}
\alpha _{ijl}{\bf e}_{Mi}{\bf e}_{Pj}{\bf e}_{Ql}=-{e^3nk_{2\perp}\omega _Q\over 2m^2\Omega^2}\left\{{\omega ^2\over\Omega ^2}\cos\phi _1
+i\epsilon {\omega\over\Omega}\sin\phi _1\right\},
\end{equation}
so with ion plasma frequency $\omega _p^2=4\pi e^2n/m$, $v_A^6\omega _p^6 = c^6\Omega ^6$, 
and $\omega << \Omega$
\begin{equation}
{dN_M\left({\bf k}\right)\over dt}={1\over 8}
\int k_{2\perp}^2{\hbar\omega _QN_Q\left({\bf k}_2\right)\over nm}
N_P\left({\bf k}_1\right)\delta ^3\left({\bf k}-{\bf k}_1-{\bf k}_2\right)\delta\left(\omega _M-\omega _P-\omega _Q\right)
\delta\left(\left|{\bf k}\right| - \left|{\bf k}_1\right|\right)\sin ^2\phi _1{d^3{\bf k_1}\over 2\pi}{d^3{\bf k_2}\over 2\pi}.
\end{equation}
Putting $\rho\delta v_{\perp}^2/2 = \int \hbar\omega _QN_Q d^3{\bf k}_2/\left(2\pi\right)^3\simeq \hbar\omega _QN_Q2\pi k_{2\perp}^2\left|k_{2||}\right|/\left(2\pi\right)^3$ we find
\begin{eqnarray}
{dN_M\left({\bf k}\right)\over dt}&=&{1\over 16}
\int {\delta v_{\perp}^2\over \left|k_{2||}\right|}
N_P\left({\bf k}_1\right)\delta ^3\left({\bf k}-{\bf k}_1-{\bf k}_2\right)\delta\left(\omega _M-\omega _P-\omega _Q\right)
\delta\left(\left|{\bf k}\right| - \left|{\bf k}_1\right|\right)\sin ^2\phi _1d^3{\bf k_1}d^3{\bf k_2}\\
&=& {1\over 32\pi }\int {\delta v_{\perp}^2\over\left|k_{||}-k_{1||}\right|v_A}N_P\left({\bf k}_1\right)
\delta\left(\left|{\bf k}\right| - \left|{\bf k}_1\right|\right)\sin ^2\phi _1d^3{\bf k_1}\\
&=& {1\over 64}{k_{1\perp}^2\delta v_{\perp}^2\over\left(1-\cos\psi\right)k_{||}v_A}N_P\left({\bf k}_1\right)
\end{eqnarray}
in the special case that $k_{\perp}=0$. This recovers the heuristic expression for $k\sim k_1$ and $1-\cos\psi=1/64$ or 
$\psi\simeq 1/4\sqrt{2}$, where $\psi$ is the angle between ${\bf k}$ and ${\bf k}_1$. With 
$N_M\propto\delta v_{\perp}^2\propto k_{\perp}^{-\alpha}$, Equation A10 gives $\alpha = 2$, and the cascade rate $\gamma\propto k_{\perp}^2\delta v_{\perp}^2$ is constant with $k_{\perp}$.


\begin{thebibliography}{}
\bibitem[Ashkin(1970)]{ashkin70}Ashkin, A. 1970, Phys. Rev. Lett., 24, 156
\bibitem[Ashkin et al.(1986)]{ashkin86}Ashkin, A., Dziedzic, J. M., Bjorkholm, J. E., \& Chu, S. 1986, Optics Letters, 11, 288
\bibitem[Asplund et al.(2009)]{asplund09} Asplund, M., Grevesse,
    N., Sauval, A.~J., \& Scott, P.\ 2009, \araa, 47, 481
\bibitem[Athay(1981)]{athay81}Athay, R. G. 1981, \apj, 249, 340 
\bibitem[Avrett \& Loeser(2008)]{avrett08}Avrett, E. H., \& Loeser, R. 2008, \apjs, 175, 229
\bibitem[Batchelor(1986)]{batchelor86}Batchelor, G. K. 1986, The Theory of Homogenous Turbulence (Cambridge: Cambridge University Press)
\bibitem[Baker et al.(2020)]{baker20}Baker, D., van Driel-Gesztelyi, L., Brooks, D. H., D\'emoulin, P., Valori, G., Long, D. M., Laming, J. M., 
To, A. S. H., \& James, A. W. 2020, \apj, 894, 35
\bibitem[Baker et al.(2019)]{baker19}Baker, D., van Driel-Gesztelyi, L., Brooks, D. H., Valori, G., James, A. W., Laming, J. M., Long, D. M., 
D\'emoulin, P., Green, L. M., Matthews, S. A., Ol\'ah, K., \& K\"ov\'ari, Z. 2019, \apj, 875, 35
\bibitem[Beresnyak(2019)]{beresnyak19}Beresnyak, A. 2019, Living Rev. Comp. Astrophys., 5, 2
\bibitem[Butler \& Dalgarno(1980)]{butler80}Butler, S. E., \& Dalgarno, A. 1980, \aap, 85, 144
\bibitem[Dahlburg et al.(2016)]{dahlburg16} Dahlburg, R. B., Laming,
    J. M., Taylor, B. D., \& Obenschain, K. 2016, \apj, 831, 160
\bibitem[Dennis et al.(2015)]{dennis15}Dennis, B. R., Phillips, K. J. H., Schwartz, R. A., Tolbert, A. K., Starr, R. D., \& Nittler, L. R. 2015, \apj, 803, 67
\bibitem[Donati \& Landstreet(2009)]{donati09}Donati, J.-F., \& Landstreet, J. D. 2009, Ann. Rev. Astron. Astrophys., 47, 333
\bibitem[Doschek et al.(2015)]{doschek15}Doschek, G. A., Warren, H. P., \& Feldman, U. 2015, \apjl, 808, L7
\bibitem[Doschek \& Warren(2016)]{doschek16}Doschek, G. A., \& Warren, H. P. 2016, \apj, 825, 36
\bibitem[Doschek \& Warren(2017)]{doschek17}Doschek, G. A., \& Warren, H. P. 2017, \apj, 844, 52
\bibitem[Drake et al.(1997)]{drake97}Drake, J. J., Laming, J. M., \& Widing, K. G. 1997, \apj, 478, 403
\bibitem[Feldman et al.(1990)]{feldman90}Feldman, U., Widing K. G., \& Lund P. A. 1990, \apjl, 364, L21 
\bibitem[Katsuda et al.(2020)]{katsuda20}Katsuda, S., Ohno, M., Mori, K., et al. 2020, \apj, 891, 126
\bibitem[Kigure et al.(2010)]{kigure10}Kigure, H., Takahashi, K., Shibata, K., Yokoyama, T., \& Nozawa, S. 2010, PASJ, 62, 993 
\bibitem[Kingdon \& Ferland(1996)]{kingdon96}Kingdon, J. B., \& Ferland, G. J. 1996, \apjs, 106, 205
\bibitem[Koyama et al.(207)]{koyama07}Koyama, K., Tsunemi, H., Dotani, T., et al. 2007, PASJ, 59, 23
\bibitem[Laming et al.(2019)]{laming19} Laming, J.~M., Vourlidas, A., Korendyke, C., et al. 2019, \apj, 879, 124
\bibitem[Laming(2017)]{laming17} Laming, J.~M.\ 2017, \apj, 844, 153
\bibitem[Laming(2015a)]{laming15} Laming, J.~M.\ 2015, Living Reviews in Solar Physics, 12, 2
\bibitem[Laming(2015b)]{laming15b}Laming, J. M. 2015, \apj, 805, 102
\bibitem[Laming(2012)]{laming12}Laming, J. M. 2012, \apj, 744, 115 
\bibitem[Laming(2009)]{laming09}Laming, J. M. 2009, \apj, 695, 954
\bibitem[Laming(2004)]{laming04} Laming, J.~M.\ 2004, \apj, 614, 1063
\bibitem[Laming \& Drake(1999)]{laming99}Laming, J. M., \& Drake, J. J. 1999, \apj, 516,324
\bibitem[Laming et al.(1996)]{laming96}Laming, J. M., Drake, J. J., \& Widing, K. G. 1996, \apj, 462, 948
\bibitem[Laming et al.(1995)]{laming95} Laming, J.~M., Drake, J.~J., \& Widing, K.~G. 1995, \apj, 443, 416
\bibitem[Landau \& Lifshitz(1987)]{landau87}Landau, L. D., \& Lifshitz, E. M. 1987, Fluid Mechanics (Oxford: Pergamon) 
\bibitem[Lee \& Parks(1983)]{lee83}Lee, N. C., \& Parks, G. K. 1983, Phys. Fluids, 26, 724
\bibitem[Le Teuff et al.(2000)]{leteuff00}Le Teuff, Y. H., Millar, T. J., \& Markwick, A. J. 2000, \aaps, 146, 157
\bibitem[Lodders(2003)]{lodders03}Lodders, K. 2003, \apj, 591, 1220
\bibitem[Lundin \& Guglielmi(2006)]{lundin06}Lundin, R., \& Guglielmi, A. 2006, SSRv, 127, 1
\bibitem[Luo \& Melrose(2006)]{luo06}Luo, Q., \& Melrose, D. 2006, \mnras, 368, 1151
\bibitem[Melrose(1986)]{melrose86}Melrose, D. B. 1986, Instabilities in Space and Laboratory Plasmas (Cambridge: Cambridge University Press) 
\bibitem[Mitsuda et al.(2007)]{mitsuda07}Mitsuda, K., Bautz, M., Inoue, H., et al. 2007, PASJ, 59, 1
\bibitem[Ng \& Bhattacharjee(1997)]{ng97}Ng, C. S., \& Bhattacharjee, A. 1997, Phys. Plasmas, 4, 606
\bibitem[Peierls(1991)]{peierls91}Peierls, R. 1991, More Surprises in Theoretical Physics (Princeton: Princeton University Press) 
\bibitem[Phillips \& Dennis(2012)]{phillips12}Phillips, K. J. H., \& Dennis, B. R. 2012, \apj, 748, 52
\bibitem[Pottasch(1963)]{pottasch63} Pottasch, S.~R.\ 1963, \apj, 137, 945
\bibitem[Reardon et al.(2008)]{reardon08}Reardon, K. P., Lepreit, F., Carbone, V., \& Vecchio, A. 2008, \apjl, 683, L207
\bibitem[Reiners et al.(2009)]{reiners09}Reiners, A., Basri, G., \& Browning, M. 2009, \apj, 692, 538
\bibitem[Ridpath(1985)]{ridpath85}Ridpath, I. 1985, A Comet Called Halley (Cambridge: Cambridge University Press)
\bibitem[Schlemm et al.(2007)]{schlemm07}Schlemm, C. E., Starr, R. D., Ho, G. C., et al. 2007, SSRv, 131, 393
\bibitem[Schwerdtfeger \& Nagle(2019)]{schwerdtfeger19}Schwerdtfeger, P., \& Nagle, J. K. 2019, Molecular Phys., 117, 1200
\bibitem[Stangalini et al.(2012)]{stangalini12}Stangalini, M., Giannattasio, F., Del Moro, D., \& Berrilli, F. 2012, \aap, 539, L4
\bibitem[Sylwester et al.(2014)]{sylwester14}Sylwester, B., Sylwester, J., Phillips, K. J. H., K\c{e}pa, A., \& Mrozek, T. 2014, \apj, 787, 122
\bibitem[Warren(2014)]{warren14}Warren, H.P. 2014, \apjl, 786, L2
\bibitem[Washimi \& Karpman(1976)]{washimi76}Washimi, H., \& Karpman, V. I. 1976, JETP, 44, 528
\bibitem[Wood et al.(2018)]{wood18}Wood, B. E, Laming, J. M., Warren, H. P., \& Poppenhager, K. 2018, \apj, 862, 66
\bibitem[Wood et al.(2012)]{wood12}Wood, B. E., Laming, J. M., \& Karovska, M. 2012, \apj, 753, 76 
\bibitem[Zhao et al.(2005)]{zhao05}Zhao, L. B., Stancil, P. C., Gu, J.-P., Liebermann, H.-P., Funke, P., Buenker, R. J., \& Kimura, M.
2005, Phys. Rev. A., 71, 062713
\end{thebibliography}
\end{document}